\documentclass[11pt]{article}
\usepackage{times}
\usepackage{setspace}
\usepackage{geometry}
\usepackage{amsmath, amsfonts}
\usepackage{graphicx, subfigure, wrapfig, url}
\usepackage{epstopdf}
\newcommand{\Prob}{\text{Prob}}

\newcommand{\prob}{\text{Prob}}

\newcommand{\qed}{\nobreak \ifvmode \relax \else
      \ifdim\lastskip<1.5em \hskip-\lastskip
      \hskip1.5em plus0em minus0.5em \fi \nobreak
      \vrule height0.75em width0.5em depth0.25em\fi}

\title{Modeling Dynamical Influence in Human Interaction\footnote{This research was sponsored 
by AFOSR under Award Number FA9550-10-1-0122 and 
ARL under Cooperative Agreement Number W911NF-09-2-0053.
The views and conclusions contained in this document are those of the
authors and should not be interpreted as representing the official
policies, either expressed or implied, of AFOSR, ARL or the U.S. Government.}}
\author
{Wei Pan$^1$, Wen Dong$^1$, Manuel Cebrian$^{1,2}$, Taemie Kim$^1$, \\ James H. Fowler$^2$, Alex (Sandy) Pentland$^1$\\
\normalsize{$^1$Media Laboratory, MIT}\\
\normalsize{20 Ames Street, Cambridge, MA 01239, USA}\\
\normalsize{$^2$University of California, San Diego}\\
\normalsize{9500 Gilman Drive, La Jolla, CA 92093, USA}\\
\normalsize{\normalsize{\{panwei, wdong, taemie, pentland\}@media.mit.edu}, \{mcebrian, jhfowler\}@ucsd.edu}
}

\begin{document}
\maketitle
\begin{abstract}
How can we model influence between individuals in a social system, even when the network of interactions is unknown?  In this article, we review the literature on the ``influence model,'' which utilizes independent time series to estimate how much the state of one actor affects the state of another actor in the system.  We extend this model to incorporate dynamical parameters that allow us to infer how influence changes over time, and we provide three examples of how this model can be applied to simulated and real data.  The results show that the model can recover known estimates of influence, it generates results that are consistent with other measures of social networks, and it allows us to uncover important shifts in the way states may be transmitted between actors at different points in time.
\end{abstract}

\section{Introduction}
The concept of \emph{influence} is extraordinarily important in the natural sciences.  The basic idea of influence is that an outcome in one entity can cause an outcome in another entity.  Flip over the first domino, and the second domino will fall.  If we understand exactly how two dominoes interact --- how one domino \emph{influences} another --- and we know the initial state of the dominoes and how they are situated relative to one another, then we can predict the outcome of the whole system. 

For decades, social scientists have also been interested in analyzing and understanding \emph{who influences whom} in social systems \cite{watts2007influentials, woolley2010evidence}.  But the analogue with the physical world is not exact.  In the social world, influence can be more complicated because internal states are often unobservable, intentional behavior can be strategic, and the situational context of specific interactions can change the effect one actor has on another.  And even more challenging, actors can choose \emph{with whom} they interact, which can confound efforts to infer influence from correlated behaviors between actors \cite{manski1993identification}.  As a consequence, there has been tremendous interest in developing methods for better understanding the effect that networked interactions have on the spread of social behaviors and outcomes.  

Social scientists have already carefully studied communication settings like group discussions to better understand the causal mechanisms that underlie influence \cite{gibson2010making}, but  recent advances in modern sensing systems such as sociometric badges~\cite{kim2008} and cell phones~\cite{eagle2006reality} now provide valuable social behavioral signals from each individual at very high resolution in time and space.  The challenge for those of us interested in signal processing is how to use this data to make better inferences about influence within social systems.

In this article we describe the ``influence model'' first articulated in \cite{basu2001learning} and the subsequent literature that has refined this approach. Similar definitions on influence in other literature include research on voting models in physics~\cite{castellano2009statistical}, cascade models in epidemiology~\cite{watts2007influentials}, attitude influence in psychology~\cite{friedkin2010attitude} and information exchange models in economics~\cite{acemoglu2010spread}.  The influence model is built on an explicit abstract definition of influence: an entity's state is influenced by its network neighbors' states and changes accordingly.  Each entity in the network has a specifically defined strength of influence over every other entity in the network, and, equivalently, each relationship can be weighted according to this strength.  

We believe that the influence model is a unique tool for social scientists because it can be applied to a wide range of social systems (including those where aggregates like organizations, states, and institutions can themselves be thought of as ``actors'' in a network).  The influence model also enables researchers to infer interactions and dynamics when the network structure is unknown---all that is needed is information about time series signals from individual observations.  And although this method is subject to the same limitations as any observational network study \cite{shalizi2010homophily}, the ordering of behaviors in time and social space makes it less likely that alternative mechanisms like selection effects and contextual heterogeneity can explain the patterns of influence ascertained by the model.

The rest of this article is organized in the following way.  We first describe the 
influence model in Section \ref{sec:influence} and previous works in Section \ref{sec:review}. 
In Section \ref{sec:dynamical} we introduce the dynamical influence 
model, a generalization of
the influence model for changing network topology.
We then discuss the inference algorithms in Section \ref{sec:inference}.
In Section \ref{sec:application}, we give specific examples of its ability to 
recover plausible and known influence pathways between entities in a network
with real and artificial data.  

\section{\label{sec:influence}Overview for the Influence Model}
\subsection{Entities in a Social System}
We describe the influence model here, followed by
a review on its history in Section \ref{sec:review}. 
The model starts with a system of $C$ entities. We assume that each entity $c$ is associated with 
a finite set of possible states ${1,\dots,S}$. At different time $t$, 
each entity $c$ 
is in one of the states, denoted by $h^{(c)}_t \in \{1,\dots,S\}$.
It is not necessary that each entity is associated with the same
set of possible states. Some entities can have more or less states. However,
to simplify our description, we assume that each entity's latent 
state space is the same without loss of generality.

The state of each entity is not directly observable. However,
as in the Hidden Markov Model (HMM), each entity emits a signal $O^{(c)}_t$ 
at time stamp $t$ based on the current latent state $h^{(c)}_t$, following a 
conditional emission probability $\prob(O^{(c)}_t | h^{(c)}_t)$. 
The emission probability can either be
multinomial or Gaussian for discrete and continuous cases respectively, 
exactly as in HMM literature~\cite{bishop2006pattern}.

It is important to note here that entities can be anything that has at least one state.  For example, they could be people in group discussions who are in a ``talking'' state or a ``silent'' state.  Or they could be geographical districts with variation in flu incidence that yields some in a ``high incidence'' state or a ``low incidence'' state.  The fundamental question remains in any situation, does the state in one entity influence (cause a change) the state in another entity?  It is therefore possible to apply the influence model to a wide range of contexts.

\subsection{Influence between Entities}
The influence model is
composed of entities 
interacting and influencing each other. 
``Influence'' is defined as
 the conditional dependence between
each entity's current state $h^{(c)}_t$ at time $t$ and the previous states of 
all entities $h^{(1)}_{t-1},\dots,h^{(C)}_{t-1}$ at time $t-1$.  
Therefore, intuitively, $h^{(c)}_t$ is 
\emph{influenced} by all other entities. 

An important implication of this Markovian assumption is that all effects from states at times earlier than $t-1$ are completely accounted for by incorporating all information from time $t-1$.  This does not mean that earlier times had no effect or were unimportant -- it just means that their total effect is felt in the immediately-previous time period.  And even path dependent processes (of which there are many in the social sciences) can operate this way, one time period at a time.

We now discuss the conditional probability:
\begin{equation}
\prob(h^{(c')}_t | h^{(1)}_{t-1},\dots,h^{(C)}_{t-1}).
\label{eq1}
\end{equation}
Once we have $\prob(h^{(c')}_t | h^{(1)}_{t-1},\dots,h^{(C)}_{t-1})$, we naturally
achieve a generative stochastic process. As in the coupled Markov Model~\cite{brand1997coupled}, 
we can take a general combinatorial 
approach Eq. \ref{eq1}, and convert this model to an equivalent 
Hidden Markov Model (HMM), in which each different
latent state combination of $(h^{(1)}_{t-1},\dots,h^{(C)}_{t-1})$ 
is represented by a 
unique state. Therefore, for a system 
with $C$ interacting entities, the equivalent
HMM will have a latent state space of size $S^C$, exponential to the number
of entities in the system, which generates insurmountable computational challenges in real applications.

The influence model,
on the other hand, uses a much simpler mixture approach with far fewer
parameters. Entities 
$1,\dots,C$ influence the state of $c'$ in the following way:
\begin{equation}
\prob(h^{(c')}_t | h^{(1)}_{t-1},\dots,h^{(C)}_{t-1}) 
= \sum_{c\in \{1,...,C\}} \underbrace{{\mathbf R}_{c',c}}_{\text{tie strength}} \times  
\underbrace{\text{Infl}(h^{(c')}_{t}| h^{(c)}_{t-1})}_{\text{influence $c \rightarrow c'$}},
\label{condprob}
\end{equation}
where ${\mathbf R}$ is a $C\times C$ matrix. (${\mathbf R}_{c_1,c_2}$ represents
the element at the $c_1$-th row and the $c_2$-th column of the matrix ${\mathbf R}$) 
${\mathbf R}$ is row stochastic, i.e., each row of this matrix sums up
to one. $\text{Infl}(h^{(c')}_{t}| h^{(c)}_{t-1})$ is modeled using a $S \times S$
row stochastic matrix ${\mathbf M}^{c,c'}$, so that 
$\text{Infl}(h^{(c')}_{t}| h^{(c)}_{t-1}) = {\mathbf M}^{c,c'}_{h^{(c)}_{t-1},h^{(c')}_t}$,
where ${\mathbf M}^{c,c'}_{h^{(c)}_{t-1},h^{(c')}_t}$ represents the element at the $h^{(c)}_{t-1}$-th
row and $h^{(c')}_{t}$-th column of matrix ${\mathbf M}^{c,c'}$. The
row stochastic matrix ${\mathbf M}^{c,c'}$ captures the influence from
$c$ over $c'$, and is very similar to 
the transition matrix in the HMM literature~\cite{bishop2006pattern}.

Eq. \ref{condprob} can be viewed as follows: all entities' states at time $t-1$
will influence the state of entity $c'$ at time 
$t$. However, the strength of influence
is different for different entities: the strength of $c$ over $c'$ 
is captured by ${\mathbf R}_{c', c}$. As
a result, the state distribution for entity $c'$
at time $t$ is a combination of influence 
from all other entities weighted
by their strength over $c'$. 
Such definition of influence from neighbor nodes
is not unique, and it has been well studied in statistical
physics and psychology 
as well\cite{castellano2009statistical,friedkin2010attitude}.
Because ${\mathbf R}$ captures influence strength between
any two entities, we refer to ${\mathbf R}$ as the \emph{Influence Matrix}.

Generally, for each entity $c$, there are $C$ different transition 
matrices in the influence model to account for the influence dynamics between
 $c$ and $c', c'=1,\dots,C$. However, it can be simplified 
by replacing the $C$ different matrices 
with only two $S\times S$ matrices ${\mathbf E}^c$ and ${\mathbf F}^c$: 
${\mathbf E}^c={\mathbf M}^{c,c}$, which captures the self-state transition; 
Empirically, in many systems an entity $c$ may influence other entities 
in the same manner. For instance, a strong politician always 
asks everyone to support his political view no matter who they are. 
Therefore, sometimes we can simplify the system by assuming  
${\mathbf M}^{c,c'}={\mathbf F}^c, \forall c' \neq c$. 

\subsection{Inference}
The influence model is a generative model defined by
parameters ${\mathbf R}, {\mathbf E}^{1:C}, {\mathbf F}^{1:C}$ and the emission
probabilities $\prob(O^{(c)}_t | h^{(c)}_t), \forall c$. 
As in most generative machine learning models, these parameters are not
set by users, but they are automatically 
learned from observations $O^{1}_{1:T},\dots,O^{C}_{1:T}$.
The inference algorithms for learning these parameters will be discussed
in Section \ref{sec:dynamical}.

The influence model has two key advantages 
over other machine learning approaches.  
First, the number of parameters grows quadratically with the latent
space size $S$ and linearly to the number of entities $C$.
As a result, the influence model 
 is resistant to overfitting when training data is limited compared with other approaches 
~\cite{dong2007modeling}.

Second, the model captures the tie strength between entities 
using a $C \times C$ matrix ${\mathbf R}$. ${\mathbf R}$ inferred
by our model can be naturally treated as the adjacency matrix 
for a directed weighted graph between nodes. This key contribution 
connects the conditional probabilistic dependence 
to a weighted network topology. In fact, the most common usage for 
the influence model in the literature is to use ${\mathbf R}$ 
to infer social structure~\cite{madan2009modeling,choudhury2004modeling} .

\section{\label{sec:review}Previous Applications of the Influence Model}
\begin{figure}[htb]
\centering
\includegraphics[width=0.80\textwidth]{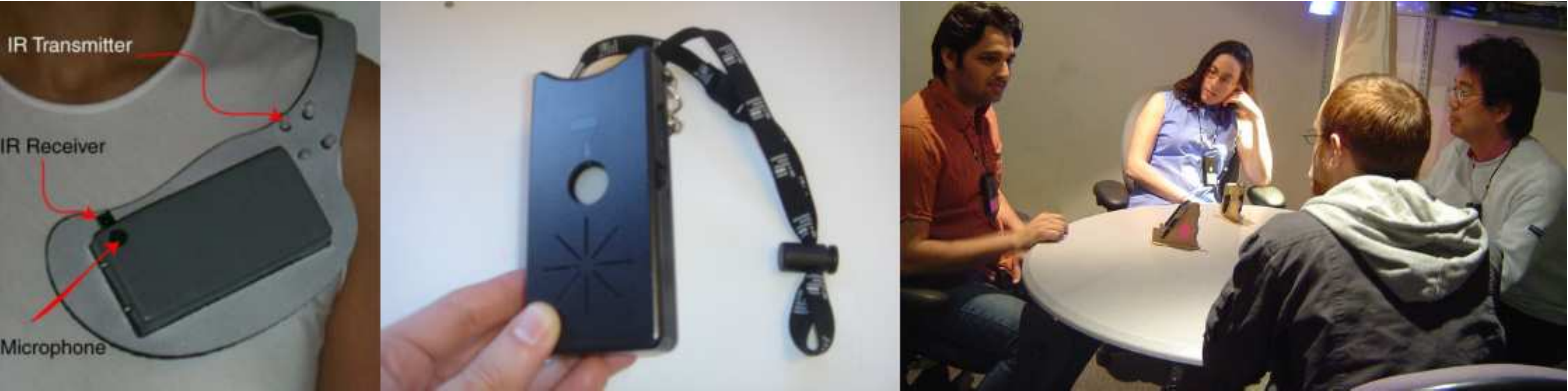}
\caption{Different versions of the 
sociometric badge is shown in the left
and in the middle. The sociometric badge 
is a wearable sensing device for
collecting individual behavioral data. 
On the right is a group brainstorming
session, and all participants were 
wearing the sociometric badges.}
\label{badgecompose}
\end{figure}

The influence model has been applied to various social systems, particularly those that have been monitored by sociometric badges like those shown in Fig. \ref{badgecompose}.  These badges are personal devices that collect individual behavioral data including audio, location, and
movement.  Early attempts to analyze data from these badges focused on questions revolving around group interaction and interpersonal influence.

The first application of the influence model \cite{basu2001learning} 
attempted to infer influence networks from audio recordings of a group discussion session with five individuals. The reseachers used audio features as observations 
$O^{(c)}_t$ and modeled the latent state space to be either ``speaking'' or ``non-speaking''.
They then used the model to infer the underlying pattern of interpersonal influence from the noisy signals measured directly from each individual and their interactions on turn taking.

An important question about these inferences relates to their validity: how do we know that the measure of influence is real?  Another set of researchers applied the influence model to conversation data from sociometric badges on 23 individuals and showed that the influence strength between individuals learned by the model correlated extremely well with
individual centrality in their social networks(with $R=0.92, p<0.0004$)
~\cite{choudhury2004modeling}.
This evidence suggests that the influence matrix defined as the
weights in the conditional dependence on states of other entities 
is an important measure for the social position of the individuals
in real interaction data.  In other words, even more abstract concepts related to influence like status or social hierarchy might be captured by the inferences of the influence model.

The model has also been applied to many other human interaction contexts \cite{madan2009modeling}. For instance, researchers have used the influence model to understand
the functional role (follower, orienteer, giver, seeker, etc) of each individual  
in the mission survival group discussion dataset\cite{dong2007using}. They
found that the inferred influence matrix helped them to achieve better
classification accuracy compared with other approaches.  The model has also been applied to the  Reality Mining\cite{eagle2006reality} cellphone sensor data.  Using information from ~80 MIT affiliates as observations and constraining the latent space
of each individual to be binary ``work'' and ``home'', researchers found that
the influence matrix learned from this data matches well with the organizational 
relationship between individuals\cite{dong2007modeling}.

Recently the influence model has been extended to a variety of systems, including traffic patterns \cite{dong2009network} and flu outbreaks \cite{pan2010modeling}.  But more importantly, there have been methodological advances that allow the model to incorporate dynamic changes in the influence matrix itself\cite{pan2010modeling}.  This new approach, the Dynamical Influence Model, is a generalization of the inference model, and is discussed in the following section.

Related approaches have utilized Bayesian networks to understand and process
social interaction time series data.
Examples include coupled HMM~\cite{brand1997coupled}, 
dynamic system trees~\cite{basu2001learning} and interacting
Markov chains~\cite{zhang2005learning}.  
The key difference between these approaches and the influence model is that the influence matrix
${\mathbf R}$ connects the real network to state dependence.
%Other relevant general multi-dimensional time series approaches such as
%LDS ~\cite{Fox:NIPS08} and the
%prototype model ~\cite{pan2009unsupervised} are not able to
%recognize the network structure and weights on edges 
%between nodes in social systems.  

The key idea of the influence model is to define influence as the state dependence
for an entity on the weighted sum of states from network neighbors. 
This idea has been extensively explored by statisical physicists\cite{castellano2009statistical},
and very recently by psychologists in modeling attitude influence\cite{friedkin2010attitude}.

\section{\label{sec:dynamical}The Dynamical Influence Model}
Above, we introduced the influence model, where the influence
strength matrix ${\mathbf R}$ remains the same for all $t$.
However, there is extensive evidence leading 
us to think that influence is indeed a dynamical 
process\cite{ansari1991survival}.
This can also be seen from many real-world experiences: Friendship
is not static;
In negotiations, your most active opponent may change
due to shifts in topics or strategies over time.
Therefore, we believe that the influence between subjects may fluctuate 
as well in many social system.

Here, we demonstrate how the influence model can be extended
to the dynamical case, and we call this generalization 
the \emph{Dynamical Influence Model}.
Instead of having one single influence strength
matrix, ${\mathbf R}$, we consider a finite set of different influence
strength matrices, $\{{\mathbf R}(1),\dots,{\mathbf R}(J)\}$, each representing
a different pattern between entities. 
$J$ is a hyperparameter set by users to define the number of different
interaction patterns. Our approach is basically a switching model, and 
we also introduce the switching latent state $r_t \in \{1,\dots,J\}, t=1,\dots,T$, 
which indicates the current
active influence matrix at time $t$. Therefore, Eq. \ref{condprob} turns
into the following:
\begin{equation}
\prob(h^{(c')}_t | h^{(1)}_{t-1},\dots,h^{(C)}_{t-1}) 
= \sum_{c\in \{1,...,C\}} {\mathbf R}(r_t)_{c',c} \times  
{\text{Infl}(h^{(c')}_{t}| h^{(c)}_{t-1})}.
\label{condprobswitching}
\end{equation}
As $r_t$ switches to different values between $1$ to $J$ at different
times $t$, the dynamics are then determined 
by different influence matrices ${\mathbf R}(r_t)$.

As shown in Section \ref{sec:application:toy}, we note that it
is very important to constrain the switching of $r_t$ for two reasons:
a) In many social systems, the change of influence patterns changes slowly and gradually.
b) A prior eliminates the probability of overfitting.
Therefore we introduce the following prior for $r_t$:
\begin{align}
r_{t+1}|r_{t} & \sim\mbox{multi}(V_{r_{t},1},\cdots,V_{r_{t},J})\label{roletran},
\end{align}
where ${\mathbf V}$ is a system parameter matrix constrained 
by another hyperparameter $p^V, p^V>=0$. The prior is shown in Eq. \ref{prior}.
\begin{eqnarray}
(V_{r_{t},1},\dots,V_{r_{t},J}) & \sim\text{Dirichlet}(10^0,10^0,\dots, 10^{p^{V}},\dots,10^0).\label{prior} \\ 
 & \phantom{\sim\text{Dirichlet}(}\stackrel{\uparrow}{1},\hspace{0.15in} \stackrel{\uparrow}{2},\hspace{0.06in} \dots, \hspace{0.07in} \stackrel{\uparrow}{r_{t}},\hspace{0.06in} \dots, \hspace{0.06in} \stackrel{\uparrow}{J} \nonumber
\end{eqnarray}
This prior provides a better control of the process $r_1,\dots,r_T$. 
When $p^V=0$, the Dirichlet prior turns to a uniform distribution. However, the higher  
$p^V$ gets, the more likely $r_{t-1}$ and $r_{t}$ will be the same.

\begin{figure}[htb]
\centering
\includegraphics[width=0.60\textwidth]{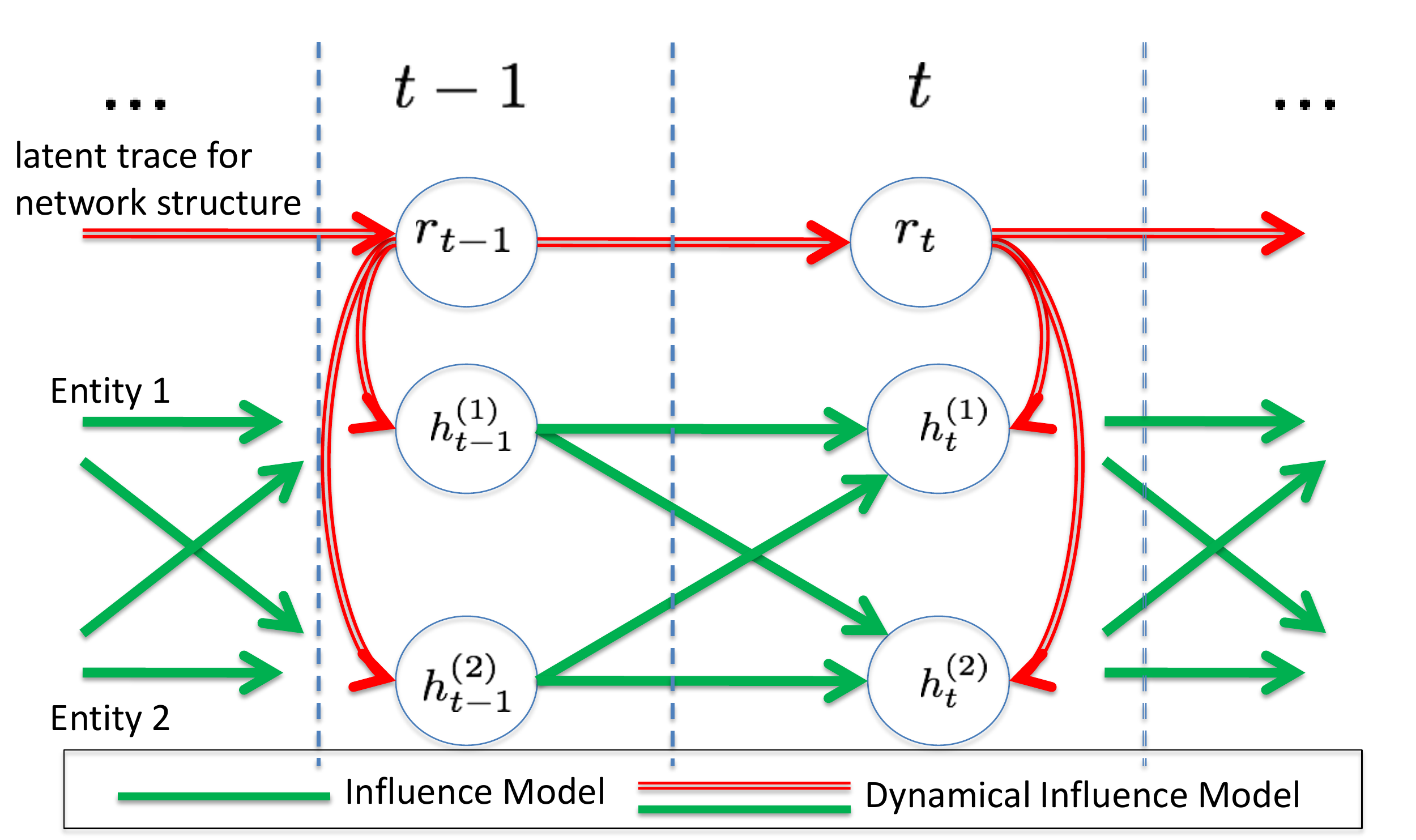}
\caption{A graphical representation of our model when $C=2$. 
The blue lines show the dependence
of the influence model described in Section \ref{sec:influence}. The red lines indicate
the layer that brings additional switching capacity to the influence model, and together
with the blue lines they fully describe the variable dependence of the dynamical influence model.}\label{figchart1}
\end{figure}

Given the model description and hyperparameters $J$ and $p^V$, we can then write
the likelihood function:\begin{eqnarray}
\lefteqn{\mathcal{L}(O^{1:C}_{1:T},h^{1:C}_{1:T},r_{1:T}|{\mathbf E}^{1:C},{\mathbf F}^{1:C} , {\mathbf R}(1:J) , {\mathbf V})}  \\ 
&=& \prod_{t=2}^{T}\Bigl\{\Prob(r_{t}|r_{t-1}) 
\times \prod_{c=1}^{C}\left[\Prob(O_{t}^{(c)}|h_{t}^{(c)}) 
 \times \Prob(h_{t}^{(c)}|h_{t-1}^{(1,...,C)},r_{t})\right]\Bigr\}  
 \nonumber \\
 &&\times \prod_{c=1}^{C}\Prob(O_{1}^{(c)}|h_{1}^{(c)})\Prob(h_{1}^{(c)})\Prob(r_{1}). 
\end{eqnarray}
To demonstrate the difference between the static influence model and the dynamical influence model, we illustrate the Bayesian graph for both models in Fig. ~\ref{figchart1}.

Researchers have been studying a variety of alternative time-varying network 
models: from EGRM~\cite{guo2007recovering} to
TESLA~\cite{ahmed2009recovering}. EGRM computes a set of features from 
networks and how they change, and models the distribution of network evolution as
the distribution of feature evolution. TESLA uses changing network edges to capture 
correlations in node observations with $l_1$ constraints on edges to ensure sparsity. 
Another recent version also learns network topology 
from cascades~\cite{gomez2010inferring}.
Compared with these models, the dynamical influence model serves as a unique generative approach
for modeling noisy signals from a dynamical network.

\section{\label{sec:inference}Inference}
In signal processing applications, we are 
given the observation time series signals $O^{(1)}_{1:T},...,O^{(C)}_{1:T}$, and
based on these observations we need to learn the distributions of 
underlying latent variables and system parameters for the dynamical
influence model. The inference process for our model is discussed here.
Since the dynamical influence model is a generalization of the
influence model, the following description is applicable to both models.

Previously, researchers started with a standard 
exact inference algorithm (Junction Tree) with exponential complexity, 
and then moved to an approach based on optimization\cite{choudhury2004modeling}.
Other scholars gradually moved to an approximation 
approach based on the Forward-Backward algorithm
and variational-EM \cite{dong2006multi,pan2010modeling}.
The influence model can also be trained via other approximations like the 
mean field method\cite{dongphd}.

Here we show some key steps for the variational E-M approach, which
has been developed and applied successfully in 
many datasets. We refer readers to 
Pan et al~\cite{pan2010modeling} for detail.
We denote definition by $\equiv$, and same distribution by $\sim$, but the right hand side of all equations should be normalized accordingly. 

{\bf E-Step:} We adopt a procedure similar to the forward-backward procedure in the HMM
literature. First, we define the following forward parameters for $t=1,...,T$.:
\begin{eqnarray}
\alpha^{r_t}_{t,c} \equiv \text{Prob}(h^{(c)}_t | r_t, O_{1:t}), \hspace{0.1in} \kappa_t \equiv \Prob(r_t|O_{1:t}),
\end{eqnarray}
where $O_{1:t}$ denotes $\{O^{(c)}_{t'}\}^{c=1,...,C}_{t'=1,...,t}$.
However, complexity for computing $\alpha^{r_t}_{t_c}$ given
$\alpha^{r_{t-1}}_{t-1,c}$ grows exponentially with respect to $C$, 
so we adopt the variational approach\cite{jordan1999introduction}, 
and E-M is still guaranteed to converge under variational 
approximation\cite{jordan1999introduction}. We proceed to decouple the chains by:
\begin{align}
\Prob(h^{(1)}_t,...,h^{(C)}_t | O_{1:t}, r_t) \approx 
\prod_c Q(h^{(c)}_t|O_{1:t}, r_t),
\end{align}
and naturally:
\begin{equation}
\alpha^{r_t}_{t,c} \approx Q(h^{(c)}_t|O_{1:t}, r_t)
\end{equation}
The approximation adopted here enables us to run inference in
polynomial time. Based on this approximation, starting with
$\alpha^{r_1}_{1,c}$ and $\kappa_1$, we can compute
$\alpha^{r_t}_{t,c}$ and $\kappa_t$, $\forall t=2,...,T$ step by step in
a forward manner.

Using the same idea, we can compute the following backward parameters
for all $t$ in the backward order (i.e. start with $t=T$, then
compute $\beta^{r_t}_{t,c}$ and $\nu_t$ for $t=T-1,T-2,...,1$):
\begin{eqnarray}
\beta^{r_t}_{t,c} \equiv \text{Prob}(h^{(c)}_t| r_t, O_{t:T}), \hspace{0.1in}  \nu_t \equiv \Prob(r_t | O_{t:T}). 
\end{eqnarray}

{\bf M-step}: With $\kappa_{t}$ and $\nu_{t}$, we can estimate:
\begin{align}
\nonumber
\lefteqn{\xi^t_{i,j} \equiv \Prob(r_t=i, r_{t+1}=j|O_{1:T}) =}\\ 
&& {\Prob(r_{t}=i|O_{1:t})
\Prob(r_{t+1}=j|O_{t+1:T})\Prob(r_{t+1}|r_t)} / \nonumber \\
&& {\sum_{i,j}\Prob(r_{t}=i|O_{1:t})
\Prob(r_{t+1}=j|O_{t+1:T})\Prob(r_{t+1}|r_t)}, \\
\lefteqn{\lambda^t_i = \Prob(r_t=i|O_{1:T}) = \frac{\sum_j \xi^t_{i,j}}{\sum_i\sum_j \xi^t_{i,j}},}
\label{rt}
\end{align}
and update $V$ by:
\begin{equation}
{\mathbf V}_{i,j} \leftarrow \frac{\sum_t \xi^t_{i,j} + k}{\sum_t \sum_j \xi^t_{i,j} + p^V},
\end{equation}
where $k=p^V$ if $i=j$, $0$ otherwise.

We then compute the joint distribution 
$\Prob(h^{q^{(c)}_{t+1}}_t, h^{(c)}_{t+1},  r_{t+1}|O_{1:T})$,
and update parameters such as influence matrices 
${\mathbf R}(1),...,{\mathbf R}(J)$, ${\mathbf E}^c$ 
and ${\mathbf F}^c$ by marginalizing this joint distribution. 

\section{\label{sec:application}Applications}
\subsection{\label{sec:application:toy}Toy Example: Two Interacting Agents}
In this example, we demonstrate how the 
dynamical influence model can be applied
to find structural changes in network dynamics.
As a tutorial, we also explain how readers should 
adjust two hyperparameters $J$ and $p^V$ in using
this model.

From a dynamical influence process composed of two interacting entities, 
we sample two binary time series of 600 steps.  
Each chain has two hidden states with a random 
transition biased to remain in the current state.
We sample binary observations from a randomly-generated multinomial distribution.
To simulate a switch in influence dynamics,
we sample with influence matrix ${\mathbf R}(1)$
(shown in Table \ref{table_configuration}) in the
first 200 frames, and later on we sample with 
influence matrix ${\mathbf R}(2)$.
We purposely make the two configuration matrices different
from each other. Partial data are shown 
in Table \ref{table_configuration} (left). 
We use the algorithm in Section \ref{sec:inference}
to infer the dynamical influence model's parameters 
${\mathbf V}, {\mathbf R}(1:J),{\mathbf E}^{1:C},{\mathbf F}^{1:C}$.
All parameters (including the emission distribution)
are initialized randomly, and they are estimated 
automatically during the E-M process.

%each chain has a higher probability to remain its own state,
%and this is to simulates the fact that human social behavior
%that one tends to persuade others to accept his own idea.

%% Not sure if we want to say this or not.
%% Table ~\ref{table_configuration}
%% show the parameters learned fully unsupervised by our algorithm
%% compared to the true values. 

\begin{table*}[htb]
\caption{Left: Part
of the two input toy sequences for a two-chain dynamical influence process. Right: The original two influence matrices of
the toy model and the same matrices learned by our algorithm with $J=3$ and $p^V=10^1$. }
\label{table_configuration}	
\begin{minipage}[b]{0.4\textwidth}
\vspace{0.5in}
\centering
\begin{tabular}{ l l}
\multicolumn{1}{c}{SEQ. NO.} &\multicolumn{1}{c}{DATA(PARTIALLY) } 
\\ \hline \\
1 & 221111121212212... \\
2 & 112111212121122... \\
\end{tabular}
\end{minipage}
\hspace{0.5cm}
\begin{minipage}[b]{0.55\textwidth}
\centering
\begin{tabular}{ l l l }
\multicolumn{1}{c}{\bf }  &\multicolumn{1}{c}{\bf ${\mathbf R}(1)$ } &\multicolumn{1}{c}{\bf${\mathbf R}(2)$} 
\\ \hline \\
 True & $\left(\begin{array}{cc}
    0.90 & 0.10  \\
    0.10 & 0.90  \\ \end{array}\right)$
 & $\left(\begin{array}{cc}
    0.05 & 0.95  \\
    0.95 & 0.05  \\ \end{array}\right)$
 \\
  Learned & $\left(\begin{array}{cc}
    0.93 & 0.07  \\
    0.10 & 0.89  \\ \end{array}\right)$
 & $\left(\begin{array}{cc}
    0.08 & 0.92  \\
    0.94 & 0.06  \\ \end{array}\right)$
 \\
\end{tabular}
\end{minipage}
\end{table*}

\begin{figure}[tb]
\centering
\subfigure[]{
\includegraphics[width=0.45\textwidth]{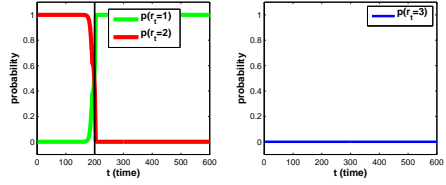}
\label{fig1}
}
\\
\subfigure[]{
\includegraphics[width=0.45\textwidth]{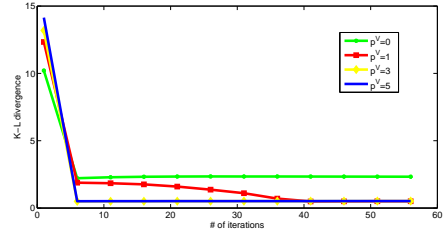}
\label{fig2}
}
\caption{(a): The posterior of $r_t$ is shown with $J=3$ after convergence. The middle black vertical line on the left indicates the true switch in $r_t$.
% (i.e. ${\mathbf R}^1$ is switched to ${\mathbf R}^2$ during sampling). 
The probability of ${\mathbf R}(1)$ being active and ${\mathbf R}(2)$ being active are shown in the left plot; ${\mathbf R}(3)$ is shown in the right, which remains inactive. 
(b): The K-L divergence between learned parameters and the true distributions with respect to number of iterations.}
\end{figure}

{\bf Choosing hyperparameters:}
We now discuss the selection of hyperparameters $J$ and $p^V$.
For the number of active influence matrices $J$, we 
illustrate their characteristics by running the
same example with $J=3$. We show the poster distribution of $r_t$
(calculated in Eq. \ref{rt}) in Fig. \ref{fig1}. 
The dynamical influence model discovers
the sudden change of influence weights accurately at $t=200$.
Since the toy process only has two true configuration matrices, 
the posterior probability of 
the $3$rd configuration being active is almost zero for 
any $t$. The system properties are fully captured
by the other two configuration matrices during the training. 
The learned configuration matrices (shown in Table \ref{table_configuration})
are correctly recovered. 
Based on Fig. \ref{fig1} and experiments with other values for
$J$ (which we cannot show here due to the space limitation),
we suggest that readers should gradually increase $J$ until
the newly added configuration matrices are no longer useful in 
capturing additional dynamical information from the data, by ensuring
there is no constant zero posterior probability as in the right plot 
in Fig. \ref{fig1}.

We also demonstrate convergence of the K-L Divergence between the
true distributions of the transition probability and
the learned distributions in Fig. \ref{fig2} with different values of $p^V$. 
As can be seen in Fig. \ref{fig2}, the algorithm converges
quickly within 50 iterations. However, when $p^V$ is small, we may
encounter over-fitting where the learned model rapidly switches
between different configurations to best suit the data.
Therefore, in Fig. \ref{fig2}, the divergence for $p^V=0$
remains higher than other $p^V$ values at convergence.
In conclusion, we advise users to increase $p^V$ gradually until
the posterior of $r_t$ does not fluctuate. 

\subsection{Modeling Dynamical Influence in Group Discussions}
\subsubsection{Dataset Description and Preprocessing}
\label{preprocess}
Researchers in \cite{kim2008} recruited 40 groups with four subjects 
in each group for this experiment. During the experiment,
each subject was required to wear a sociometric badge 
on their necks for 
audio recording, illustrated in the right picture in 
Fig. \ref{badgecompose}, and each group was required to 
perform two different group discussion tasks: a brainstorming 
task (referred as \emph{BS}) and a problem solving task (referred as \emph{PS}).
Each task usually lasted for 3 to 10 minutes. 
We refer readers to the original paper\cite{kim2008} 
for details on data collection and experiment preparations. 

The groups were asked to perform these tasks in two
different settings: one in which people were \emph{co-located} in the same 
room around a table (referred as \emph{CO}), and one in which two pairs of people 
were placed in two separate rooms with only audio 
communication available between them (referred as \emph{DS}). 
The badges are deployed in both cases for audio collecting.
We separated all samples according to their
tasks (BS/PS) and their settings (CO/DS), and we ended up
with four categories: DS+BS, DS+PS, CO+BS, CO+PS.
Since discussions were held in 
four-person groups, each sample for a discussion session
is composed of four sequences 
collected by the four badges on participants' chests.
The audio sequence picked up by each badge was split
into one-second blocks. Variances of speech energy were
calculated for each block. We then applied a 
hard threshold to convert them into binary sequences.
In all experiments, we only used binary sequences 
as data input.

\subsubsection{Predicting Turn Taking in Discussion}
We here explain an application of 
the dynamical influence model to predict turn taking, and we show
that it is possible to achieve good accuracy in prediction given only
the audio volume variance observations, with no information
from the audio content. 

Ten occurrences of turn taking behavior from each sample are
selected for prediction purposes. ``Turn taking'' here is defined
as incidences in which the current speaker ceases speaking, and 
another speaker starts to speak. 

For the dynamical influence model, we model each person
as an entity $c$, and the observed audio variances at time $t$ as
$O^{(c)}_t$. Each person also has two hidden states, representing
speaking or not speaking. The hidden layer
eliminates error due to noise and non-voicing speaking in audio
signals\cite{choudhury2004modeling}. Therefore, influence here is set to
capture how each person's speaking/non-speaking
hidden states dynamically change other people's speaking/non-speaking
states (i.e., how people influence each others' turn taking).
All parameters are initialized randomly and learned 
by the E-M inference algorithm in this example.
We train the dynamical influence model using data up to $t-1$,
sample observations at time $t$ from it, and mark the chain that
changes the most toward the high-variance observations as the turn taker at $t$.
The emission probability $\Prob(O^{(c)}_t | h^{(c)}_t)$ is modeled using a
multinomial distribution, and is estimated automatically during the E-M process.

For comparison, we also show results using TESLA and nearest neighbors methods.
For TESLA, we use the official implementation\cite{ahmed2009recovering} 
to obtain the changing weights between pairs of nodes, and we pick the node that has the strongest correlation weight to other nodes at $t-1$ as the turn taker at $t$.
To predict the turn taking at time $t$ using the nearest neighbor
method, we look over all previous instances of turn taking behaviors 
that have the same speaker as the one in $t-1$, and predict by using the most frequent outcomes.

%% \begin{table*}[tb]
%% \caption{The description for four different categories of all the samples.}
%% \label{catg}
%% \begin{center}
%% \begin{tabular}{ll}
%% \multicolumn{1}{c}{\bf CATEGORY}  &\multicolumn{1}{c}{\bf TASK DESCRIPTION}
%% \\ \hline 
%% CO+PS         & Four people perform a problem solving task in the same room. \\
%% CO+BS         & Four people perform a brainstorming session in the same room. \\
%% DS+PS         & Four people perform the same problem solving task in two rooms with Skype.  \\
%% DS+BS         & Four people perform the same brainstorming session in two rooms with Skype.
%% \end{tabular}
%% \end{center}
%% \end{table*}

\begin{table*}[htb] 
\caption{Accuracy for different turn taking 
prediction methods on both the full dataset
and the half of the dataset with more complex interactions. 
The random guess accuracy is $33\%$. Human accuracy is typically
around $50\%$ for similar tasks\cite{schlangen2006reaction}. }
\label{terror}
\vspace{0.1in}
\begin{tabular}{l|llll|llll}
\multicolumn{1}{l|}{\bf } & \multicolumn{4}{|l|}{\bf ACCURACY} & \multicolumn{4}{|l}{\bf ACCURACY}
\\
\multicolumn{1}{l|}{\bf } & \multicolumn{4}{|l|}{\bf ALL SAMPLES} & \multicolumn{4}{|l}{\bf COMPLEX INTERACTION SAMPLES}
\\ \hline
{\bf METHODS} & {\bf DS+BS} & {\bf DS+PS} &{\bf CO+BS} & \multicolumn{1}{l|}{\bf CO+PS} & {\bf DS+BS} & {\bf DS+PS} &{\bf CO+BS} & {\bf CO+PS} 
\\ \hline 
%Random   &0.33&0.33&0.33&0.33&0.33&0.33&0.33&0.33
%\\
TESLA    &0.41&0.42&0.32&0.25&{0.44}&0.37&0.37&0.17
\\
NN       &{\bf 0.58}&0.60&0.48&0.50&{\bf 0.47}&0.47&0.38&0.26
\\
Ours(J=1)&0.45&{\bf 0.67}&{\bf 0.75}&{\bf 0.63}&{0.45}&0.56&{\bf 0.77}&{0.62}
\\
Ours(J=2)&0.46&0.58&0.65&0.34    &{\bf 0.47}&0.58&{0.67}&0.46
\\
Ours(J=3)&0.50&0.60&0.55&0.48          &{\bf 0.47}&{\bf 0.73}&0.65&{\bf 0.65} 
\\
\end{tabular}
\end{table*}
The accuracy for each algorithm is listed in Table \ref{terror}. We also 
show the prediction accuracy for the half of
all samples that have more complex interactions, i.e., higher entropy.
For the dynamical influence approach, we list error rates for $J=1$ (which is simply
the influence model), $J=2$ and $J=3$.
Except DS+BS, We notice that the dynamical influence model
outperforms others in all categories with different $J$. 
This performance is quite good considering that 
we are using only volume and that
a human can only predict at around $50\%$ accuracy for
similar tasks\cite{schlangen2006reaction}.

More importantly, the dynamical influence model
 seems to perform much better than the competing methods for more
complex interactions. For simple interactions, it seems 
that $J=1$ or even NN perform
the best due to the fact that there is little shift in the 
influence structure during
the discussion. However, when handling complex 
interaction processes, the introduction
of a switching influence dynamic dramatically improves 
the performance as shown in Table \ref{terror}.
This result suggests that the dynamical influence assumption 
is reasonable and necessary in modeling complex group dynamics, 
and it can improve 
prediction accuracy significantly. However, 
in simple cases, the model achieves 
the highest performance when $J=1$ (i.e. the 
influence matrix is static), and a higher $J$ will only lead to overfitting.

The fact that turn taking is predictable using our 
dynamical influence assumption is indeed surprising,
because group turn taking dynamics 
are complicated and related to content as well~\cite{gibson2010making}.
We think that the dynamical influence model tracks
the two main mechanisms of group discussions noted in 
Gibson~\cite{gibson2010making} even though the model does not incorporate the content of speaker statements.  First, the time variant assumption in the 
dynamical influence model captures the latency factor in 
group dynamics.  And second, our abstract conditional probability definition
of influence is essentially a generalization of the conversational obligation mechanism Gibson described.

\subsection{Modeling Flu Epidemics as Influence Dynamics}
\begin{figure}[h]
\centering
\includegraphics[width=0.30\textwidth]{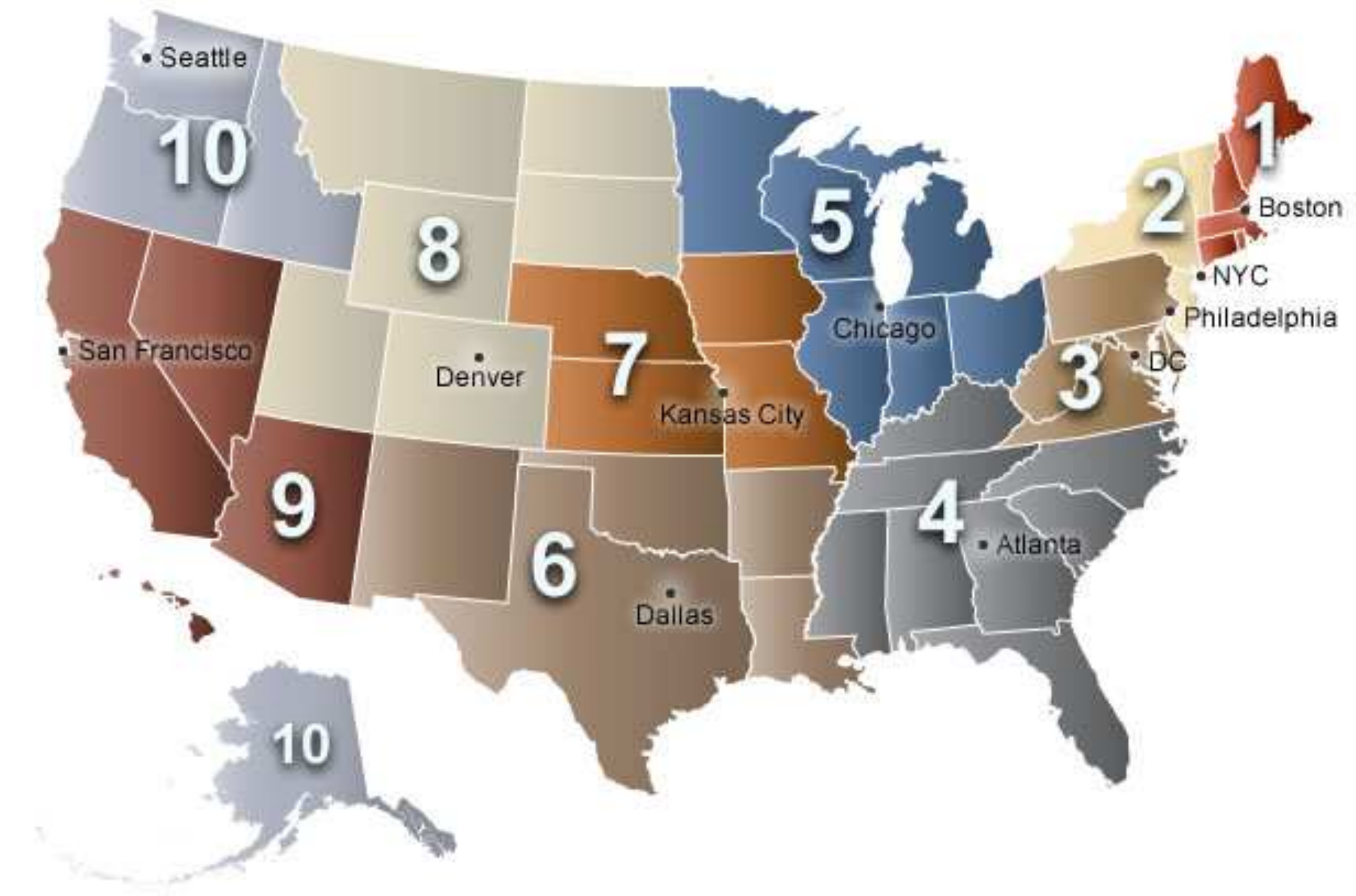}
\caption{Ten regions of the United States defined by US Health and Human Services.}
\label{hhs}
\end{figure}

The last example is an application of the dynamical influence model to weekly US flu activity data from Google Flu Trends \cite{ginsberg2008detecting}. All 50 states
in the U.S. are divided into ten regions by their geo-location, as shown in Fig. \ref{hhs}, and we model each region as an entity in the dynamical
influence model.

\begin{figure*}[htb]
\centering
\includegraphics[width=0.95\textwidth]{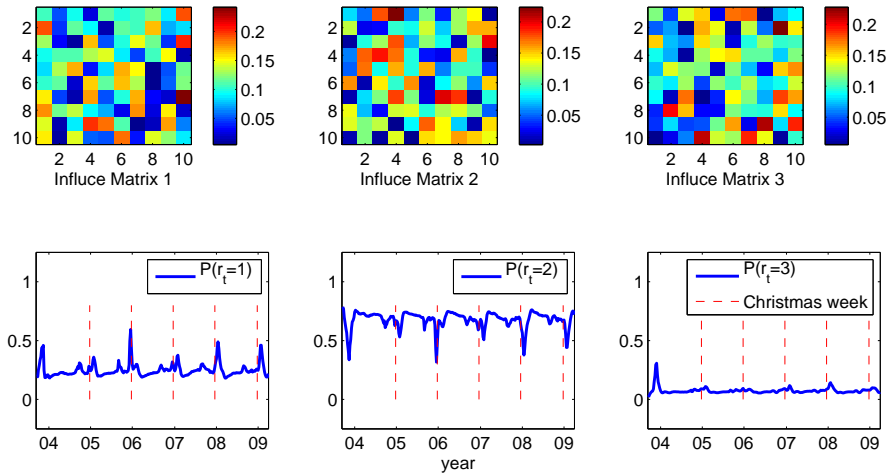}
\caption{The inferred posterior for $r_t$ given all observations
after convergence is shown here. 
While there are many small peaks indicating changes in
influence, the largest peaks occur at Christmas holiday seasons,
which implies holiday traffic patterns can have a big effect on flu transmissibility between regions.  We find that three configuration matrices 
are good enough to capture the flu dynamics. }\label{fig3}
\end{figure*}

As the data are continuous, 
six hidden states are used for each chain, 
and $p(O_t^{(c)}|h_t^{(c)})$ is modeled with six different
Gaussian distributions with different means and the same variance for each hidden state. We set by hand
the six mean values so that they 
represent the six different severe levels
for the flu epidemics, from the least severe to the most severe.
We train the model using the first 290 weeks (from 2003 to early 2009), 
and we show the posterior for $r_t$, the switching parameter,
in Fig.\ref{fig3} together
with the three learned influence matrices.
While there are many small peaks suggesting changes
in influence, the probability changes dramatically
around Christmas, which 
suggests that the influence patterns among these
ten regions are very different during the 
holiday season.  Note especially that we did not tell the model to search for different patterns on those days -- instead, it reveals the fact that transmission dynamics operate differently during a time when many people are engaging in once-a-year travel to visit family and friends.  While it is possible that other mechanisms are at work, alternative explanations like the weather are not plausible because they do not change discontinuously during the holiday season. 

Influence matrix $1$ captures the dynamics during holiday seasons,
while influence matrix $2$ captures the dynamics during normal seasons. 
Row $i$ corresponds to the region $i$ in Fig. \ref{hhs}. Let's take
an example of Row 1, the New England region. During normal times as shown
in the 1st row of influence matrix $2$, New England is more likely to
be influenced by close regions such as $3$ and $4$; during holiday
seasons, New England is more likely to be influenced by all regions
especially distant regions such as region $9$.  The same phenomena exist
for other regions as well.
\section{Discussion}
In this article we described the influence model and its generalization, the dynamical
influence model, and we showed how these can be applied to a variety of social signals to infer how entities in networks affect one other. 
In particular, we can use the resulting influence matrix ${\mathbf R}$ 
to connect the underlying network and the stochastic process of
state transition. The switching matrices 
${\mathbf R}(1),...,{\mathbf R}(J)$ are even able to
bridge the state transition to time-varying networks. 

The influence model shares the same issues with other machine learning models: inference requires sufficient training data, and tuning is necessary for best results. Future work includes combining known network data into the model to boost performance.  

The most important limitation is that we are attempting to infer causal processes from observational data in which many mechanisms are likely at play.  If we find that behavior between two individuals is correlated, it could be due to influence, but it could also be due to selection (I choose to interact with people like me) or to contextual factors (you and I are both influenced by an event or a third party not in the data).  It has been recently shown that these mechanisms are generically confounded \cite{shalizi2010homophily} but it is important to remember that this does not make observational data worthless.  It just means that we should carefully consider alternative mechanisms that may underlie correlated behavior.  The fact that we have time data to test causal ordering and we have asymmetries in network relationships to test direction of effects means that we can have greater (but not complete!) confidence than we would if we only had cross-sectional data from symmetric relationships.
%But regardless of the underlying mechanisms, the influence model performs very well in predicting behavior, both in simulation studies and in out-of-sample tests from real data.
%In future work we hope to apply the model to a wide range of settings to infer hidden networks and status hierarchies.  We also hope to leverage known network data to boost the performance of the model when data is limited.

\bibliographystyle{unsrt}
{\footnotesize
\bibliography{panwei}

\begin{thebibliography}{10}

\bibitem{watts2007influentials}
D.J. Watts and P.S. Dodds.
\newblock {Influentials, networks, and public opinion formation}.
\newblock {\em Journal of Consumer Research}, 34(4):441--458, 2007.

\bibitem{woolley2010evidence}
A.W. Woolley, C.F. Chabris, A.~Pentland, N.~Hashmi, and T.W. Malone.
\newblock {Evidence for a collective intelligence factor in the performance of
  human groups}.
\newblock {\em science}, 330(6004):686, 2010.

\bibitem{manski1993identification}
C.F. Manski.
\newblock Identification of endogenous social effects: The reflection problem.
\newblock {\em The Review of Economic Studies}, 60(3):531, 1993.

\bibitem{gibson2010making}
D.R. Gibson.
\newblock Making the turn.
\newblock {\em Social Psychology Quarterly}, 73(2):132, 2010.

\bibitem{kim2008}
T.~Kim, A.~Chang, L.~Holland, and A.S. Pentland.
\newblock {Meeting mediator: enhancing group collaborationusing sociometric
  feedback}.
\newblock In {\em Proceedings of the ACM 2008 conference on Computer supported
  cooperative work}, pages 457--466. ACM, 2008.

\bibitem{eagle2006reality}
N.~Eagle and A.~Pentland.
\newblock {Reality mining: sensing complex social systems}.
\newblock {\em Personal and Ubiquitous Computing}, 10(4):255--268, 2006.

\bibitem{basu2001learning}
S.~Basu, T.~Choudhury, B.~Clarkson, A.~Pentland, et~al.
\newblock {Learning human interactions with the influence model}.
\newblock {\em MIT Media Laboratory Technical Note}, 2001.

\bibitem{castellano2009statistical}
C.~Castellano and V.~Loreto.
\newblock {Statistical physics of social dynamics}.
\newblock {\em Reviews of Modern Physics}, 81(2):591, 2009.

\bibitem{friedkin2010attitude}
N.E. Friedkin.
\newblock The attitude-behavior linkage in behavioral cascades.
\newblock {\em Social Psychology Quarterly}, 73(2):196, 2010.

\bibitem{acemoglu2010spread}
D.~Acemoglu, A.~Ozdaglar, and A.~ParandehGheibi.
\newblock Spread of (mis) information in social networks.
\newblock {\em Games and Economic Behavior}, 2010.

\bibitem{shalizi2010homophily}
C.R. Shalizi and A.C. Thomas.
\newblock Homophily and contagion are generically confounded in observational
  social network studies.
\newblock {\em Arxiv preprint arXiv:1004.4704}, 2010.

\bibitem{bishop2006pattern}
C.M. Bishop et~al.
\newblock {\em {Pattern recognition and machine learning}}.
\newblock Springer New York:, 2006.

\bibitem{brand1997coupled}
M.~Brand, N.~Oliver, and A.~Pentland.
\newblock {Coupled hidden Markov models for complex action recognition}.
\newblock In {\em IEEE Computer Society Conference on Computer Vision and
  Pattern Recognition}, pages 994--999, 1997.

\bibitem{dong2007modeling}
W.~Dong and A.~Pentland.
\newblock {Modeling influence between experts}.
\newblock {\em Lecture Notes in Computer Science}, 4451:170, 2007.

\bibitem{madan2009modeling}
A.~Madan and A.~Pentland.
\newblock {Modeling Social Diffusion Phenomena using Reality Mining}.
\newblock In {\em AAAI Spring Symposium on Human Behavior Modeling. Palo Alto,
  CA}, 2009.

\bibitem{choudhury2004modeling}
T.~Choudhury and S.~Basu.
\newblock {Modeling conversational dynamics as a mixed memory markov process}.
\newblock In {\em Proc. of Intl. Conference on Neural Information and
  Processing Systems (NIPS)}. Citeseer, 2004.

\bibitem{dong2007using}
W.~Dong, B.~Lepri, A.~Cappelletti, A.S. Pentland, F.~Pianesi, and M.~Zancanaro.
\newblock {Using the influence model to recognize functional roles in
  meetings}.
\newblock In {\em Proceedings of the 9th international conference on Multimodal
  interfaces}, pages 271--278. ACM, 2007.

\bibitem{dong2009network}
W.~Dong and A.~Pentland.
\newblock {A Network Analysis of Road Traffic with Vehicle Tracking Data}.
\newblock 2009.

\bibitem{pan2010modeling}
W.~Pan, M.~Cebrian, W.~Dong, T.~Kim, and A.~Pentland.
\newblock {Modeling Dynamical Influence in Human Interaction Patterns}.
\newblock {\em Arxiv preprint arXiv:1009.0240}, 2010.

\bibitem{zhang2005learning}
D.~Zhang, D.~Gatica-Perez, S.~Bengio, and D.~Roy.
\newblock {Learning Influence Among Interacting Markov Chains}.
\newblock 2005.

\bibitem{ansari1991survival}
S.A. Ansari, V.S. Springthorpe, and S.A. Sattar.
\newblock {Survival and vehicular spread of human rotaviruses: possible
  relation to seasonality of outbreaks}.
\newblock {\em Reviews of infectious diseases}, 13(3):448--461, 1991.

\bibitem{guo2007recovering}
F.~Guo, S.~Hanneke, W.~Fu, and E.P. Xing.
\newblock {Recovering temporally rewiring networks: A model-based approach}.
\newblock In {\em Proceedings of the 24th international conference on Machine
  learning}, page 328. ACM, 2007.

\bibitem{ahmed2009recovering}
A.~Ahmed and E.P. Xing.
\newblock {Recovering time-varying networks of dependencies in social and
  biological studies}.
\newblock {\em Proceedings of the National Academy of Sciences}, 106(29):11878,
  2009.

\bibitem{gomez2010inferring}
M.~Gomez~Rodriguez, J.~Leskovec, and A.~Krause.
\newblock {Inferring networks of diffusion and influence}.
\newblock In {\em Proceedings of the 16th ACM SIGKDD international conference
  on Knowledge discovery and data mining}, pages 1019--1028. ACM, 2010.

\bibitem{dong2006multi}
W.~Dong and A.~Pentland.
\newblock {Multi-sensor data fusion using the influence model}.
\newblock In {\em Proceedings of the International Workshop on Wearable and
  Implantable Body Sensor Networks}, page~75. Citeseer, 2006.

\bibitem{dongphd}
{Dong, Wen}.
\newblock {\em {Modeling the Structure of Collective Intelligence}}.
\newblock PhD thesis, {Massachusetts Institute of Technology}, {2010}.

\bibitem{jordan1999introduction}
M.I. Jordan, Z.~Ghahramani, T.S. Jaakkola, and L.K. Saul.
\newblock {An introduction to variational methods for graphical models}.
\newblock {\em Machine learning}, 37(2):183--233, 1999.

\bibitem{schlangen2006reaction}
D.~Schlangen.
\newblock {From reaction to prediction: Experiments with computational models
  of turn-taking}.
\newblock In {\em Ninth International Conference on Spoken Language
  Processing}. Citeseer, 2006.

\bibitem{ginsberg2008detecting}
J.~Ginsberg, M.H. Mohebbi, R.S. Patel, L.~Brammer, M.S. Smolinski, and
  L.~Brilliant.
\newblock {Detecting influenza epidemics using search engine query data}.
\newblock {\em Nature}, 457(7232):1012--1014, 2008.

\bibitem{weiss2009variational}
R.J. Weiss and D.P.W. Ellis.
\newblock {A variational EM algorithm for learning eigenvoice parameters in
  mixed signals}.
\newblock In {\em Proceedings of the 2009 IEEE International Conference on
  Acoustics, Speech and Signal Processing-Volume 00}, pages 113--116. IEEE
  Computer Society, 2009.

\bibitem{chang2001libsvm}
C.C. Chang and C.J. Lin.
\newblock {LIBSVM: a library for support vector machines}, 2001.

\bibitem{dimicco2007}
J.M. DiMicco, KJ~Hollenbach, A.~Pandolfo, and W.~Bender.
\newblock {The Impact of Increased Awareness while Face-to-Face}.
\newblock {\em Human-Computer Interaction}, 22(1), 2007.

\end{thebibliography}
}

\newpage

\section{Appendix A: Model Learning}
We here show detail steps for our variational E-M algorithm. 
Definition is denoted by $\equiv$, and $\sim$ denotes
the same distribution but the right side should be normalized
accordingly. 
\subsection{E-Step} 
We adopt a procedure similar to the forward-backward procedure in HMM
literature. We compute the following forward parameters for $t=1,...,T$.:
\begin{eqnarray}
\alpha^{r_t}_{t,c} \equiv \text{Prob}(h^{(c)}_t | r_t, O_{1:t}), \\
\kappa_t \equiv \Prob(r_t|O_{1:t}),
\end{eqnarray}
where $O_{1:t}$ denotes $\{O^{(c)}_{t'}\}^{c=1,...,C}_{t'=1,...,t}$.
However, exact inference is not intractable. We apply the variational approach
in ~\cite{weiss2009variational}~\cite{jordan1999introduction}. The variational E-M process is still guaranteed to converge because of the lower bound property of the variational method~\cite{jordan1999introduction}.
We decouple the chains by:
\begin{align}
\Prob(h^{(1)}_t,...,h^{(C)}_t | O_{1:t}, r_t) \approx 
\prod_c Q(h^{(c)}_t|O_{1:t}, r_t),
\end{align}
and naturally:
\begin{equation}
\alpha^{r_t}_{t,c} \approx Q(h^{(c)}_t|O_{1:t}, r_t)
\end{equation}

We define: 
\begin{equation}
\alpha_{t,c} \equiv \Prob(h^{(c)}_t | O_{1:t}) = \sum_{r_t} \kappa_t \alpha^{r_t}_{t,c},
\end{equation}

%% %%%%%%%%%Those are wrong historical stuff
%% %We now define: 
%% %\begin{eqnarray}
%% %\label{defineh}
%% %\lefteqn{H^{r_t} \equiv} \\ 
%% %\nonumber
%% %&&\left( \begin{array}{cccc}
%% %{\mathbf R}^{r_t}_{1,1}T^{(1)} & {\mathbf R}^{r_t}_{1,2}F^{(2)} & \cdots & {\mathbf R}^{r_t}_{1,C}F^{(C)} \\
%% %{\mathbf R}^{r_t}_{2,1}F^{(1)} & {\mathbf R}^{r_t}_{2,2}T^{(2)} & \cdots & {\mathbf R}^{r_t}_{2,C}F^{(C)} \\
%% %\vdots & \vdots & \ddots & \vdots \\
%% %{\mathbf R}^{r_t}_{C,C}F^{(1)} & \cdots & \cdots & {\mathbf R}^{r_t}_{C,C}T^{(C)}
%% %\end{array} \right).
%% %\end{eqnarray}

%% %In the following content, the notation $\overrightarrow{}$ represents
%% %the concatenation of the discrete distribution vector. Moreover,
%% %$\overrightarrow{\alpha}^{r_t}_t$ denotes the concatenation of 
%% %the discrete distribution vectors of $\alpha^{r_t}_{t,1},...,\alpha^{r_t}_{t,C}$, and 
%% %${\overrightarrow{O_t}}$ denotes the concatenation of the distribution vectors
%% %of $\Prob(h^{(1)}_t|O^{(1)}_t), ..., \Prob(h^{(C)}_t|O^{(C)}_t)$.
%% %%%%%%%%%End of wrong historical stuff

and
\begin{eqnarray}
\widehat{{\alpha}^{r_t}}_{t-1,c} \equiv
\Prob(h^{(c)}_{t-1}|r_t, O_{1:t-1}) = \sum_{r_{t-1}} \widehat{\kappa^{r_t}}_{t-1} {\alpha}^{r_{t-1}}_{t-1,c}, 
\end{eqnarray}
where:
\begin{eqnarray}
\widehat{\kappa^{r_t}}_{t-1} &=& \Prob(r_{t-1}|O_{1:t-1}, r_t).
%\\ 
%&\sim&
%{\Prob(r_{t-1}|O_{1:t-1}) \Prob(r_{t}|r_{t-1})}.
%{\sum_{r_t}\Prob(r_t|O_{1:t}) \times \Prob(r_{t+1}|r_t)}
\end{eqnarray}
We define $Q(h^{(c)})$ to be in the form of $\frac{\Prob(h^{(c)}_t |O_{1:t-1}, r_t) s^{(c)}_t}{\sum_{h^{(c)}_t} \Prob(h^{(c)}_t |O_{1:t-1}, r_t) s^{(c)}_t}$, which captures both the evidence from previous states($\Prob(h^{(c)}_t |O_{1:t-1}, r_t)$) and the evidence($s^{(c)}_t$) from observations. We then have:
\begin{eqnarray}
\nonumber
\lefteqn{Q(h^{(1,...,C)}_t| O_{1:t}, r_t)} \\
&& \equiv \prod_c \frac{ 
\underbrace{\left(\sum_{h^{(c)}_{t-1}}{\mathbf R}^{r_t}_{c,c} {\mathbf E}^{(c)}_{h^{(c)}_{t-1},h^{(c)}_t}\widehat{\alpha^{r_t}}_{t-1,c} 
+ \sum_{c',c' \neq  c}\sum_{h^{(c')}_{t-1}}{\mathbf R}^{r_t}_{c,c'} {\mathbf F}^{(c')}_{h^{(c')}_{t-1}, h^{(c)}_t} \widehat{\alpha^{r_t}}_{t-1, c'}
\right)}_{\Psi^{(c)}} \times s^{(c)}_t}
{\sum_{h^{(c)}_t} \left(
\sum_{h^{(c)}_{t-1}}{\mathbf R}^{r_t}_{c,c} {\mathbf E}^{(c)}_{h^{(c)}_{t-1},h^{(c)}_t}\widehat{\alpha^{r_t}}_{t-1,c} 
+ \sum_{c',c' \neq  c}\sum_{h^{(c')}_{t-1}}{\mathbf R}^{r_t}_{c,c'} {\mathbf F}^{(c')}_{h^{(c')}_{t-1}, h^{(c)}_t}\widehat{\alpha^{r_t}}_{t-1, c'} 
\right) \times s^{(c)}_t
},
\label{forward}
\end{eqnarray}
where $\Psi^{(c)}$ is actually $\Prob(h^{(c)}_t|O_{1:t-1}, r_t)$. We also have:
\begin{align}
\nonumber
\lefteqn{\Prob(h^{(1,...,C)}_t, O_t| O_{1:t-1}, r_t) =} \\ 
&&\prod_c \underbrace{ \left(
\sum_{h^{(c)}_{t-1}}{\mathbf R}^{r_t}_{c,c} {\mathbf E}^{(c)}_{h^{(c)}_{t-1},h^{(c)}_t} \widehat{\alpha^{r_t}}_{t-1,c}
+ \sum_{c',c' \neq  c}\sum_{h^{(c')}_{t-1}}{\mathbf R}^{r_t}_{c,c'} {\mathbf F}^{(c')}_{h^{(c')}_{t-1}, h^{(c)}_t}
\widehat{\alpha^{r_t}}_{t-1, c'}
\right)}_{\Psi^{(c)}} \Prob(O^{(c)}_t | h^{(c)}_t),
\end{align}
and 
\begin{align}
\Prob(h^{(1,...,C)}_t | O_{1:t-1}, r_t) = \frac{\Prob(h^{(1,...,C)}_t, O_t|O_{1:t-1}, r_t)}{\Prob(O_t | O_{1:t-1}, r_t)}.
\end{align}

We continue to minimize the K-L divergence between $\Prob(h_t^{(1,...,C)}|O_{1:t}, r_t)$
and $Q(h_t^{(1,...,C)}|O_{1:t}, r_t)$, that is:
\begin{eqnarray}
\nonumber
\lefteqn{\arg\min_{s^{(c)}_t}{\mathbb D} \equiv {\mathbb E}_Q \left(\log Q(h_t^{(1,...,C)}|O_{1:t}, r_t)\right) 
- {\mathbb E}_Q \left( \Prob(h^{(1,...,C)}_t|O_{1:t}, r_t) \right) }\\
\nonumber
&&={\mathbb E}_Q\left(\sum_c \log \Psi^{(c)} + \sum_c \log s^{(c)}_t - \sum_c \log (\sum_{h^{(c)}_t}\Psi^{(c)} s^{(c)}_t) \right) \\
&& \hspace{0.2in}  - {\mathbb E}_Q\left(\sum_c \log \Psi^{(c)} + \sum_c \log \Prob(O^{(t)}|h^{(c)}_t)\right) + \underbrace{\Prob(O_t | O_{1:t-1}, r_t)}_{\text{unrelated to }s^{(c)}_t }. 
\end{eqnarray}
By taking the derivative we have:
\begin{align}
\nonumber
\lefteqn{\frac{\partial {\mathbb D}}{\partial s^{(c)}_t}= 
\sum_{h^{(c)}_t}\frac{\partial \widehat{\alpha^{r_t}}_{t,c}}{\partial s^{(c)}_t}
\left(s^{(c)}_t - \Prob(O^{(c)}_t|h^{(c)}_t) \right) = 0}  \\
&& \Rightarrow s^{(c)}_t =  \Prob(O^{(c)}_t|h^{(c)}_t)
\end{align}
We then compute $\kappa_t$ using Bayes' rule:
\begin{align}
\kappa_t  \sim \Prob(O_t|r_t,O_{1:t-1})\Prob(r_t|O_{1:t-1}). 
\label{kappaforward}
% \\ \nonumber
%&& \prod_c\left(\sum_{h^{(c)}_t} \Prob(O_t|h_t) \Prob(h_t|r_t, O_1:O_{t-1}) \right) \\
%&& \times \Prob(r_t|O_1,O_{t-1})
%&& \times \Prob(r_t|r_{t-1})
\end{align}
where $\Prob(O_t|r_t,O_{1:t-1})$ can be evaluated using the previous
approximation results. The prior part of Eq. \ref{kappaforward}
can be evaluated using ${\mathbf V}$ and $\kappa_{t-1}$.

Using the same idea, we can compute the following backward parameters
for all $t$:
\begin{eqnarray}
\beta^{r_t}_{t,c} \equiv \text{Prob}(h^{(c)}_t| r_t, O_{t:T}), \\ 
\nu_t \equiv \Prob(r_t | O_{t:T}), \\
\beta_{t,c} \equiv \Prob(h^{(c)}_t | O_{t:T}) = \sum_{r_t} \nu_t \beta^{r_t}_{t,c}.
\end{eqnarray}

%%%%%%%%% Detail omitted

% By calculating
% \begin{eqnarray}
% \widehat{\nu^{r_t}}_{t+1} &\equiv& \Prob(r_{t+1}|O_{t+1:T}, r_t)\\
% \nonumber 
% &\sim& {\Prob(r_{t+1}|O_{t+1:T})  \Prob(r_{t}|r_{t+1})},
% %%
% %% \frac sum {\sum_{r_{t+1}}\Prob(r_{t+1}|O_{t+1}:O_T) \times \Prob(r_{t}|r_{t+1})} 
% \end{eqnarray}
% and
% \begin{eqnarray}
% \widehat{{\beta}^{r_t}}_{t+1,c} &\equiv&
% \Prob(h^{(c)}_{t+1}|r_t, O_{t+1:T}) \\
% &=& \sum_{r_{t+1}}  \widehat{\nu^{r_t}}_{t+1} \beta^{r_{t+1}}_{t,c},
% \end{eqnarray}
% we can easily establish the proposition for the backward procedure as
% the forward procedure described above.

% %%%%%%%%%%%%%%
% \begin{proposition}
% \begin{eqnarray}
% \overrightarrow{\beta}^{r_t}_{t} = \sum_{r_{t+1}}\overrightarrow{O_t}^T H^{r_{t+1}} \otimes 
% \widehat{\overrightarrow{\beta}^{r_{t+1}}}_{t+1},
% \end{eqnarray}
% where:
% \begin{eqnarray}
% \widehat{{\beta}^{r_t}}_{t-1,c} &\equiv&
% \Prob(h^{(c)}_{t+1}|r_t, O_{t+1:T}). 
% \end{eqnarray}
% \end{proposition}
% %%%%%%%%%%%%%%

% Similar to Eq. \ref{kappaforward}, we have:
% \begin{align}
% \nonumber
% \lefteqn{{\nu}_{t} \sim} \\
% &&\Prob(O_t|r_t, O_{t+1:T}) \Prob(r_t| O_{t+1:T}).
% % omitting some details
% % && \prod_c \left(\sum_{h^{(c)}_t} \Prob(O_t|h^{(c)}_t) \Prob(h_t|r_t, O_{t+1}:O_T) \right) \\
% % &&\times\Prob(r_t|r_{t+1}) \\
% \end{align}

%%%%%%%%%%%%%%%%% END OF DETAIL

% The auxiliary variables calculated below is used in the M-step and
% experiment demonstrations.

\subsection{M-step}
With $\kappa_{t}$ and $\nu_{t}$, we can estimate:
\begin{align}
\nonumber
\lefteqn{\xi^t_{i,j} \equiv \Prob(r_t=i, r_{t+1}=j|O_{1:T}) =}\\ 
&& \frac{\Prob(r_{t}=i|O_{1:t})
\Prob(r_{t+1}=j|O_{t+1:T})\Prob(r_{t+1}|r_t)}{\sum_{i,j}\Prob(r_{t}=i|O_{1:t})
\Prob(r_{t+1}=j|O_{t+1:T})\Prob(r_{t+1}|r_t)},
\end{align}
and
\begin{eqnarray}
\lambda^t_i &=& \Prob(r_t=i|O_{1:T}) = \frac{\sum_j \xi^t_{i,j}}{\sum_i\sum_j \xi^t_{i,j}}.
\label{rt}
\end{eqnarray}
We then update $V$ by:
\begin{equation}
{\mathbf V}_{i,j} \leftarrow \frac{\sum_t \xi^t_{i,j} + k}{\sum_t \sum_j \xi^t_{i,j} + p^V},
\end{equation}
where $k=p^V$ if $i=j$, $0$ otherwise.

We compute the following joint distribution.
\begin{eqnarray}
\label{comb1}
\lefteqn{\Prob(h^{q^{(c)}_{t+1}}_t, h^{(c)}_{t+1}, q^{(c)}_{t+1}, r_{t+1}|O_{1:T}) = }
\\ 
\nonumber
&& \begin{cases} \frac{1}{Z}
T^{(c)}_{h^{(c)}_t, h^{(c)}_{t+1}} \times  \widehat{\alpha^{r_t}}_{t, c} \beta^{r_t}_{t+1,c} \lambda^t \Prob(q^{(c)}_{t+1}|r_{t+1}) , \\
\hspace{1.2in}\text{if $q^{(c)}_{t+1}=c$,} \\
\frac{1}{Z}F^{q^{(c)}_{t+1}}_{h^{q^{(c)}_{t+1}}_t, h^{(c)}_{t+1}} \times \widehat{\alpha^{r_t}}_{t,q^{(c)}_{t+1}}\beta^{r_t}_{t+1,c} \lambda^t \Prob(q^{(c)}_{t+1}|r_{t+1}) , \\
\hspace{1.2in}\text{if $q^{(c)}_{t+1} \neq c$.}
\end{cases}
\end{eqnarray}

$Z$ denotes the normalization factor that can be calculated easily by summing all possible
values for each variable. This is fast to compute since the joint distribution is made of only four variables.
By marginalizing Eq. ~\ref{comb1}, we can update parameters ${\mathbf
  R}, {\mathbf E}$ and ${\mathbf F}$:

\begin{equation}
{\bf R}^j_{c_1,c_2} \leftarrow \frac{\sum_t\Prob(q^{(c_1)}_ t = c_2, r_t=j |O_{1:T} )}{\sum_t \sum_c \Prob(q^{(c_1)}_t=c, r_t=j|O_{1:T})},
\end{equation}

\begin{equation}
{\mathbf E}^{(c)}_{s_i,s_j} \leftarrow \frac{\sum_t \Prob(h^{(c)}_t=s_i, h^{(c)}_{t+1}=s_j, q^{(c)}_t=c | O_{1:T})}
                       {\sum_t\sum_{s} \Prob(h^{(c)}_t=s_i, h^{(c)}_{t+1}=s, q^{(c)}_t=c | O_{1:T})},
\end{equation}
and
\begin{align}
\nonumber
\lefteqn{{\mathbf F}^{(c)}_{s_i,s_j}} \\
&&\leftarrow \frac{\sum_t \sum_{c'} \Prob(h^{(c)}_t=s_i, h^{(c')}_{t+1}=s_j, q^{(c')}_{t+1}=c|O_{1:T})}
                       {\sum_t \sum_{c'} \sum_s \Prob(h^{(c)}_t=s_i, h^{(c')}_{t+1}=s, q^{(c')}_{t+1}=c|O_{1:T})}.
\end{align}

\newpage
\section{Appendix B: Detecting Structural Changes in the Discussion Dynamics}
We here provide an additional example for detecting structural changes using
the same dataset described in Section 5 of our paper.

One important feature of this model is its ability 
to capture changes in influence dynamics given
only observed time series for each node. In 
this section, we will demonstrate the performance
of our model in detecting changes with the group 
discussion dataset. 

In our discussion, \emph{a sample} refers to the set
of four sequences collected by the four badges in deployed in a group
discussion session.
We adopt the following evaluation procedure: One mixed binary audio 
sample for each four-person group is generated by concatenating  
the co-located discussion session sample and the the distributed discussion 
session sample of the same group. 
It is known that \cite{kim2008} the interaction pattern in 
a distributed discussion session is often different from a 
co-located discussion session.
Therefore, we are able to create ground truth
about changes of influence patterns
by switching from a distributed discussion sample to a co-located discussion
sample manually.
It should be noted that we only use binary sequences by thresholding the volume
variance. Thus, we have eliminated all information in the 
audio content. Two samples from each group are included in our 
final evaluation set: a) the original sample of the 
co-located discussion session (CO) and 
b) the mixed sample as described above (CO+DS). 
We end up with a total of 28 groups and 56 samples
in the final set. Lengths of each sample vary
 from 100 seconds to 500 seconds.

We apply our model on both samples for each group.
The emission probability in our model is used to
tolerate possible error due to hard thresholding and possible noise. 
We choose $J=2$, and $p^V$ is optimized for best 
performance. The posterior of $r^t$ for the two samples from 
each group is stored as the output of 
the algorithm. 

We continue to develop 
simple heuristics for distinguishing DS+CO
from CO by looking at the difference 
of the expected influence matrix  
($\sum_j \lambda^t_j R_j$) at $t=1$ 
and $t=0.8T$ for each sample, and
the one with larger difference is labeled as the CO+DS sequence. 
Given the pair of samples for each group, we test the labeling accuracy 
based on the output of our model. 
For comparison, we also implement two 
other techniques: a) classification based on one single 
feature, the turn taking rate, and 
b) S.V.M.-based classification (using implementation
in \cite{chang2001libsvm}). It is well recognized that
the turn taking rate is an important indicator for 
group dynamics. We compute the two turn taking
rates for each pair of samples and compare them 
to determine sample labels. For S.V.M., 
we compute the turn taking rate and the speaking durations for
each group member as the feature vector for each sample. Its performance is 
obtained via a four-fold cross validation. 
It should be emphasized that the S.V.M. classification task 
is different from the other two, and it is naturally
more challenging: all samples are mixed together before 
fed to S.V.M. rather than being fed to
other two algorithms in a pairwise manner.

We must point out that the 
ground truth in our evaluation may not be 
accurate: There is no guarantee in the 
dataset that a group of people 
behave and interact with each other 
differently when they are performing discussions 
using remote communication tools rather 
than being in the same room.

\begin{figure}[tb] 
\centering
\includegraphics[width=0.60\textwidth]{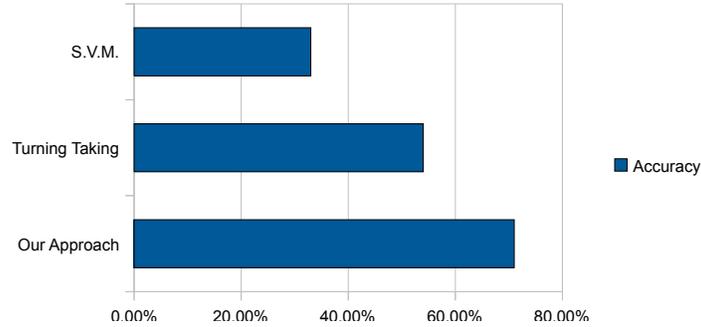}
\caption{The accuracy rates for classifying CO+DS samples from CO samples are
shown above. Our algorithm performs significantly better than
the other two methods, which are based on simple statistical features.}
\label{labac}
\end{figure}

We illustrate the accuracy rates in 
Fig. ~\ref{labac}. As we expected, our algorithm reaches $71\%$
accuracy and outperforms the other two methods. 
We argue that the influence dynamic is 
an intrinsic property of the group, 
which can not be fully revealed using simple 
statistical analysis on observable features. 
To investigate and visualize the dynamical 
characteristics of human interaction patterns,
a more sophisticated model, such as our dynamical 
influence process, must be deployed to
reveal the subtle differences in influence dynamics.   

%These results strongly indicate that our algorithm outperforms other statistical approaches
In addition, we claim that our model is capable of 
modeling, quantifying and tracking occurrences 
of such shifts in face-to-face dynamics accurately.
Our model fits 
its parameters to best suit switches between 
different influence patterns, and the parameters will be helpful for
sociologists to objectively investigate the 
micro relationship in a group discussion session. 
Information discovered by
our algorithm will also be useful in applications such as understanding possible 
interventions in human interactions\cite{kim2008}\cite{dimicco2007}.

\end{document}